\documentclass[10pt,a4paper]{article} 

\usepackage{anysize}
\usepackage[format = hang]{caption}
\usepackage{amsmath,amsthm,verbatim,amssymb,amsfonts,amscd, graphicx}
\usepackage{graphics,natbib}
\usepackage{titlesec,footmisc}
\usepackage{hyperref} 

\usepackage{listings}
\usepackage{booktabs} 

\theoremstyle{plain}

\theoremstyle{definition}

\def\hatt{\widehat}

\def\beq{\begin{eqnarray}}
\def\eeq{\end{eqnarray}}

\def\beqn{\begin{eqnarray*}}  
\def\eeqn{\end{eqnarray*}}

\def\hatt{\widehat}

\titleformat{\section}{\normalfont\large\sc\centering}{\thesection}{1em}{}
\titleformat{\subsection}[runin]{\normalfont\large\bfseries}{\thesubsection}{1em}{}
\numberwithin{equation}{section} 
\renewenvironment{abstract}
               {\list{}{\rightmargin\leftmargin}%
                \item[\text{\hspace{10mm}\sc Abstract.}]\relax}
               {\endlist}

\begin{document}

\def\heute{February 25, 2026: three days after the Olympics}

\begingroup
\begin{centering} 

\Large{\bf The Best Metal-Grabbing Games Ever: \\
    How a Tiny Nation Won the Most Medals (By Far) }\\[0.8em]
\large{\bf Nils Lid Hjort} \\[0.3em] 
\small {\sc Department of Mathematics, University of Oslo} \\[0.3em]
\small {\sc {\heute}}\par
\end{centering}
\endgroup


\begin{abstract}
  \small{For three Winter Olympics in a row, tiny nation Norway
    has out-medalled everyone else, in 2026 winning 18 golds,
    12 silvers, 11 bronzes, i.e.~41 medals, compared
    to e.g.~12 + 12 + 9 = 33 for the USA, 10 + 6 + 14 = 30
    for home team Italy, 8 + 10 + 7 = 26 for powerhouse Germany, etc.
    Never before have we [pluralis proudiensis] or anyone else won
    as many as 41 medals at a Winter Olympics. But how impressive
    is this, really, when we factor in that the number of events
    has increased so drastically?
     
\noindent
{\it Key words:}
binomial process, 
gold and silver and bronze, 
national pride,
tiny nation,
tiny secrets, 
Winter Olympics
  }
\end{abstract}


\section*{Yes, let's watch the Olympics} 

\begin{figure}[h]
\centering
\includegraphics[scale=0.35]{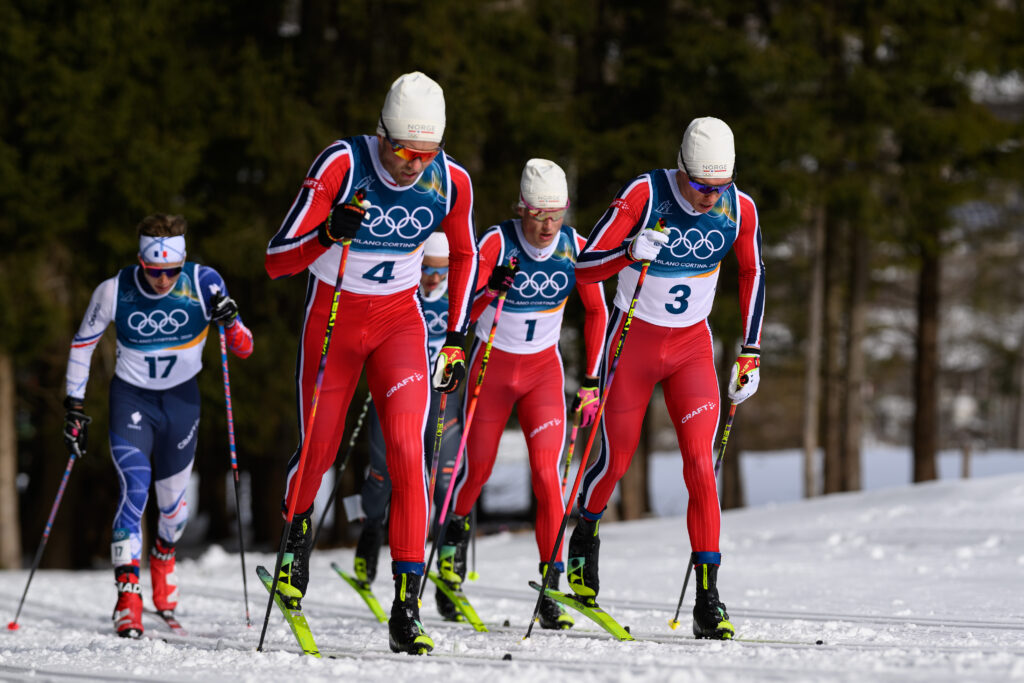}
\caption{
  Norway at the front: Klæbo, Nyenget, Iversen,
  in the 50 km cross country race. Everyone else
  (in the 2026 Games) are lagging behind.}
\label{figure:femmila}
\end{figure}


\noindent 
Indeed, the haul of metal and medals is pretty impressive, and it
has demanded attention and forces for many of us, day after day,
catching the most salient moments, living through the drama.
For these strata of people this is not merely a nationalistic
pastime, swimming in perceived passport-related glory,
but a complex sociological process, where we also follow
long lists of warm-hearted good-looking strong-willed
competitors from other countries, with attention, admiration,
and good wishes. In the Olympic Process of things
I have even been on tv and on radio and on Titan to explain
the inner-outer intricacies of speedskating to The People,
and to dicuss various other Olympic-sized Issues.

Apparently, millions {\it\&} millions followed not merely the
drama of their favourite events, but also the {\it medal table},
from day to day, ending here; see Appendix A below for
a fuller table, including correlated information using
the classic {\it Olympic points} 7-5-4-3-2-1 scoring system.
Some followers narrowly insist on
`no, gold and only gold is what matters to human beings'
(bless them; though I hesitate to bless the subgroup
who also claims `and this is the only correct way, all others are wrong'). 
There is no official UN-IOC given way of summarising
and ranking such metallic counting;
I belong to the more generous segment who look with
admiration also on the silvers and bronzes
(and to the sure fjerdeplassen). So our friends
NED have a very decent 9th place. 

{{\baselineskip12.0pt
\begin{center}
\begin{verbatim} 
                          gold silver bronze total    
                   1 NOR	  18    12    11     41   
                   2 USA	  12    12     9     33 
                   3 ITA	  10     6    14     30  
                   4 GER	   8    10     8     26  
                   5 JPN	   5     7    12     24  
                   6 FRA	   8     9     6     23  
                   7 SUI	   6     9     8     23   
                   8 CAN	   5     7     9     21   
                   9 NED	  10     7     3     20  
                  10 SWE	   8     6     4     18   
\end{verbatim}
\end{center}
}}

\section*{Sorting Metal and Counting Medals}

Even though many of us have ecumenic and altruistic
views, on our sofas, when watching the Olympics, part of
the reason for getting up in the morning to watch
the gruelling fifty-km cross-country race, minute by minute,
remains, admittedly, hoping that Our Own do well.
And, indeed, we [pluralis familiaris immodestiae et superbiae]
did exceedingly well: 18 + 12 + 11 = 41 is a new Olympic Record
(the previous being Norway 14 + 14 + 11 = 39 in PeyongChang, 2018).

But there's an element of apples and pears here, or, rather,
smaller and medium-sized and bigger apples.
The number of events has increased drastically, from 16 in Chamonix,
1924, to 34 in Innsbruck, 1964, to a high 61 at
Lillehammer and Hamar, 1994, to a whooping de lux
102, 109, 116 at PyeongChang, Beijing, Milano. 
Surely, Thorleif Haug and Johan Grøttumsbråten and Oskar Olsen
and Clas Thunberg (``Alone Against All of Norway'')
and Julius Skutnabb and Sonia Henie didn't quite see it coming,
the unheavenly legion of new types of events,
from snowboarding and halfpiping and moguls and
team figure skating and massastarts (yes, that's Dutch)
and mixed relays and autocorrelated biathlon events
(if you're a medal contender in one event,
you can win five medals) 
and big-air to curling for couples (married or not).
Hence, winning 41 medals in 116 events (in 2026)
is arguably less impressive than winning 17 medals in 16 events (in 1924).

How hard it is to win depends also, not surprisingly,
on the volume and quality of your competitors. Relevant
here, for analysing {\it\&} interpreting the 2026 results,
is the absence of Olympic powerhouse Russia (and Belarus).
One may speculate about, and even form statistical models
with fine-tuned parameters, to predict how many fewer
Norwegian medals there would have been, in the alternative
and happier counterfactual universe with the athletes
also from these nations being invited. People are people,
incidentally, regardless of their passports.
Reading the Olympic rhetoric of 1896 and 1908 it'd be fair
to say that sanctioning segments of athletes from
taking part, on the basis of where the dice of fate
have let them be born, goes against the Olympic spirit. 

\begin{figure}[h]
\centering
\includegraphics[scale=0.60]{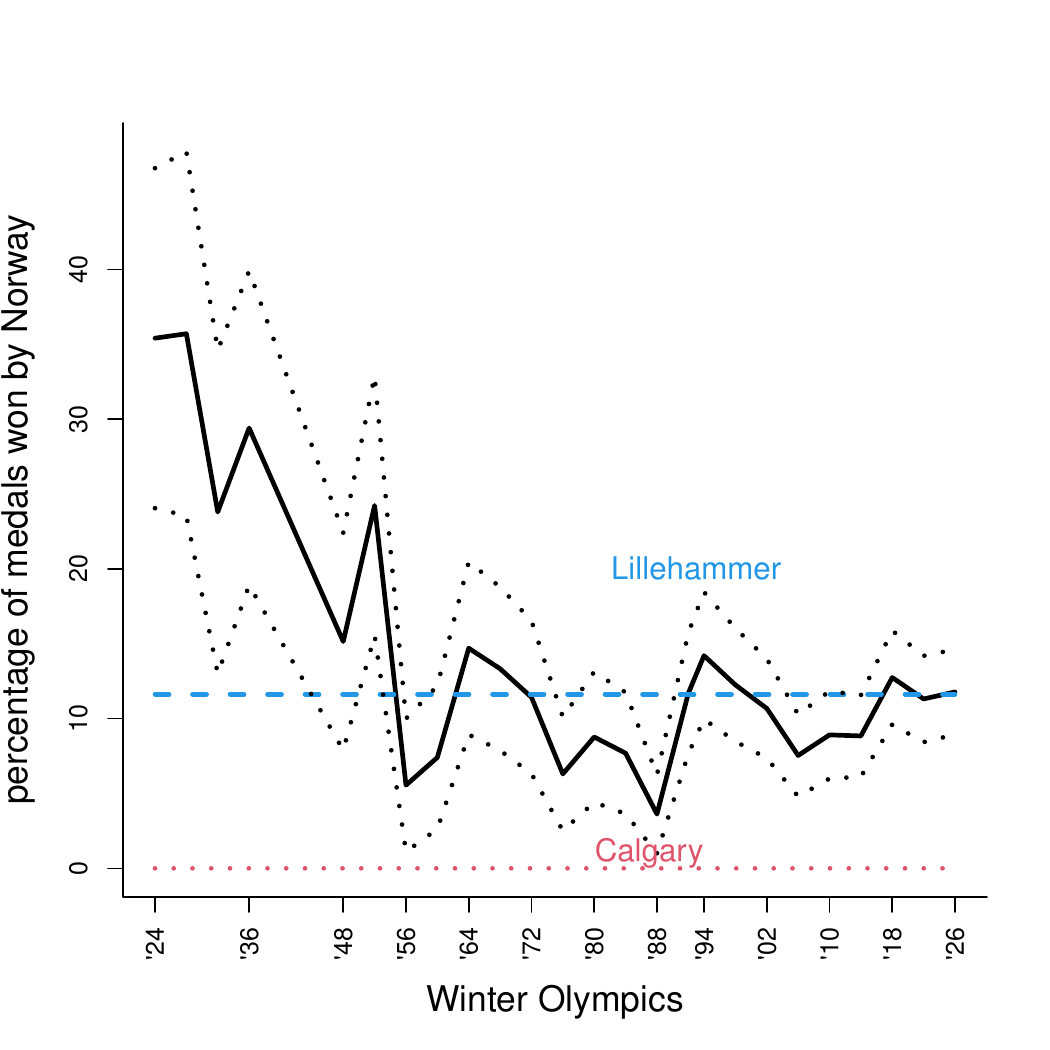}
\caption{
  Medals won by Norway, in percent (black curve),
  through Winter Olympics history, 1924 to 2026.
  The horizontal line is for the post-1960 average 11.6 \%,
  and the dotted lines indicate a 90 \% confidence band.}
\label{figure:medalsratioA}
\end{figure}

At any rate, a statistician can at least start counting,
and putting up ratios; perhaps he or she can also model
the outcomes to learn about patterns and trends and
influential factors (and predict the future).
The counting business is in terms of gold + silver + bronze
medals, and the relevant ratios, from the current Norwegian
perspective, are the percentages of Norwegian medals,
from Olympics to Olympics. Thus, in 2026 we won 18 + 12 + 11 = 41
medals out of $116\cdot3 = 348$ medal chances (11.6 \%),
which is less than winning 10 + 11 + 5 = 26 medals out of $61\cdot3 = 183$
medal chances (14.2 \%) in Lillehammer 1994.
It is similarly less than 14 medals from 37 events (13.3 \%, Grenoble, 1968)
and 15 medals of 34 events (14.7 \%, Innsbruck, 1964).
And, amazingly, Norway won 35.4 \%, 35.7 \%, 23.8 \%, 29.4 \%
of the medals in the first four Winter Olympics, as well
as about a quarter of the medals on home ground in 1952;
see the table of Appendix A. So perhaps our 41 medals
flown home from Italy isn't quite a national, or an Olympic,
or a world record, after all.

\section*{A Multinomial Process Approach to Medal Counting}

In addition to staring transfixedly at the medals, and their
numbers, which are being used for national political reasons
and thrown into various debates (should Norway ever bother
to host the Olympics again, for example, or is it easiest
for us to come to lots of other parties, grab most of the medals
and eat most of the food, party hard, and fly home),
a statistician can sit back for a moment and contemplate
the underlying processes, to check for trends, patterns,
regime shifts, and assess the variability of podia.
Here we may start by considering $\hatt p=y/(3m)$, 
for each Winter Olympics, the binomial proportion of $y$
medals won by Norway out of the $3m$ medal chances over
$m$ events. I've plotted these percentages in
Figure \ref{figure:medalsratioA}, along with 90 \% confidence intervals.
Since the very start in Chamonix 1924,
the tiny nation has won 447 medals across 1279 events,
i.e.~the astounding 34.9 \% of all medals.
The horizontal blue line in the figure is however
the more relevant 11.6 \%, the post-1960 average. 



We learn that the 2026 national data point is sufficiently
fabulous, by all means, but also that it isn't a Wondrous
Outlier, compared to the level achieved and kept in reasonable
balance since Kuppern Moe Maier Innsbruck 1964. Reading
from the band of confidence we are reminded of the
Collective National Trauma of Calgary 1988
(``og s\aa{} faller Geir Karlstad'') and the Somewhat Painful Torino 2006
(Northug sitting at home). Largely speaking, we're not in
``alt lykkes for denne regjering'' terrain;
Norway seems able to catch around 10 \% of all medals,
regardless of the colour of the  government and the
increasing number of pseudo-youth-ish Olympic events
and even the number of skiing days at Bjørnholt.
``Det er typisk norsk å være god'', as explained
Gro Harlem Brundtland to the world (well, in 1992). 


\section*{Norway, USA, Italy are Almost Equally Wondrous Countries}

Chance plays a role in these games, of course (which might be
why they're called games). Even Norwegians might admit that
Sturla Lægreid crossing the finishing line 0.2 seconds
sooner than Émilien Jacquelin in the biathlon men sprint
(enough to give him bronze and three television minutes
and world wide fame,
``I wish to be a role model, but I have failed'')
could have turned out the other way, and even
the Dutch might nod to the thought that Wiklund's 0.06 loss
to Rijpma-De Jong in the 1500 m is close to zero.
Similarly, there were as usual a little string of dramatic
half-traumatic fourth-places which very easily could have
been bronzes.

\begin{figure}[h]
\centering
\includegraphics[scale=0.60]{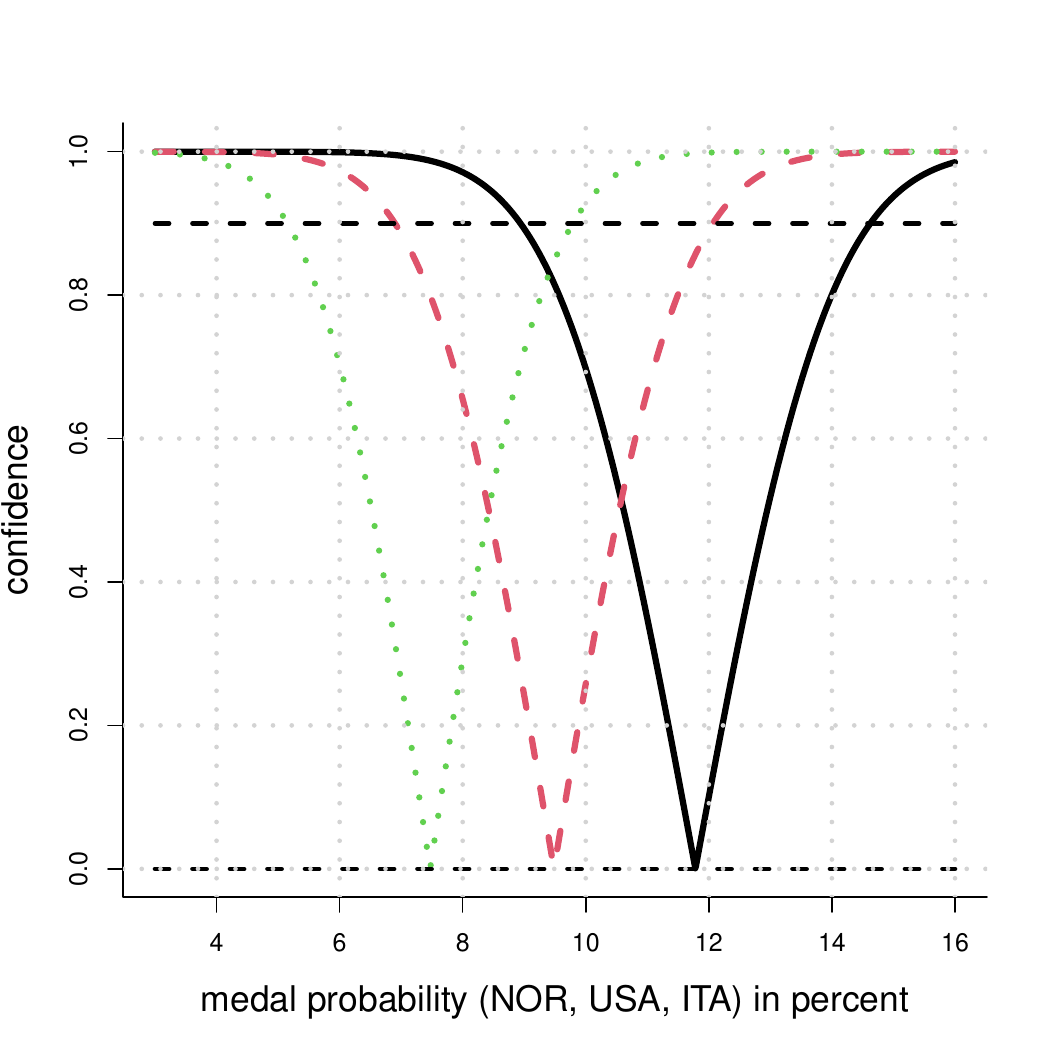}
\caption{
  Confidence curves for the three medal winning
  probabilities, for Norway, for Sambandsstatene, for Italy,
  with point estimates 11.8 \%, 9.5 \%, 7.5 \%.
  90 percent confidence intervals can be read off
  via the horizontal line. We learn that the three
  countries aren't that different, after all,
  as confidence intervals overlap.}
\label{figure:medalsratioB}
\end{figure}

With this in mind we may have another look at
the overall medal table, attempting to interpret the counts
as statistical sightings of underlying hard-core probabilities,
and then ask whether the probabilities
\beqn
p_{\rm NOR}, \quad
p_{\rm USA}, \quad
p_{\rm ITA},
\eeqn 
estimated at 
\beqn
\hatt p_{\rm NOR}={41\over 3\cdot116}=11.8\,\%,\quad 
\hatt p_{\rm USA}={33\over 3\cdot116}=9.5\,\%,\quad 
\hatt p_{\rm ITA}={26\over 3\cdot116}=7.5\,\%,
\eeqn 
are really that different.
Figure \ref{figure:medalsratioB} gives
confidence curves for the three probabilities
(via methodology of Schweder and Hjort, 2016, Ch.~3).
So 90 \% confidence intervals overlap, and there
is no significant evidence that the three countries
NOR, USA, ITA are really different, at heart.
A standard log-likelihood ratio test for the hypothesis
that the three probabilities are equal has
a non-significant p-value of 0.153.



\section*{The Tiny Country's Tiny Secrets}

Amazingly, even Big Media in other countries are writing
about the Norwegian medal catching abilities, offering
tentative explanations and speculative factors behind
the heist (and, interestingly, with most of the headlines
telling their readers that the articles give them
The Norwegian Secrets, which means we need to develop
New Secrets for the coming Olympics). You may check
The Guardian's Norse Code theories (egalitarian society,
the team experience, no jerks), TIME's crushing theories
(working hard, let the children play, bosses not being
gorillas, never asking a fellow human being how much
she or he weighs, apparently), CNN's yet further secrets
(winning by focusing on not focusing on winning,
which sounds difficult, but we manage), etc.

Yes, there's something there, in these Revealed Secrets,
which are of course not secrets but just plain facts of
Norwegian habits, instincts, lives. There are a couple of
other dimensions to point to. Yes, we still manage to be
a decently egalitarian society, where we might bump into
the king or the finance minister or Martine Ripsrud while
skiing or in the bookshop and say `hei' and we don't weaponise
our teachers and students. After all, we belong to
{\it The Almost Nearly Perfect People} (see Booth, 2014).
These (remaining) egalitarian
aspects are also reflected in how most of us look at sports.
In various other societies, we learn by our anthropological
tourist studies, there's somehow a bigger sociological
distance between the top athletes and the mainstream people,
and between the mainstreamers and say the top academics,
than in Norway. Reading Schwanitz and his fascinating
{\it Bildung. Alles, was man wissen muß},
for example, teaches us what we're supposed to know,
to have a good Bildung in life -- but also, intriguingly,
we're being told what we should rather not know
(``Was man nicht wissen sollte''). We should rather {\it not}
know about the personal records of Dutch and Soviet speedskaters
from the 1970ies and 1980ies, and neither about
Eggen vs. Tiainen and Vedenin, or Alsgaard vs.~Teichmann,
or Kulizhnikov vs.~Wotherspoon (I take it), since that's
conceivably about as bad a public stain on your personality
as paying too much attention to sladreblader
and die Regenbogepresse and paparazzi reports.

\begin{figure}[h]
\centering
\includegraphics[scale=0.17]{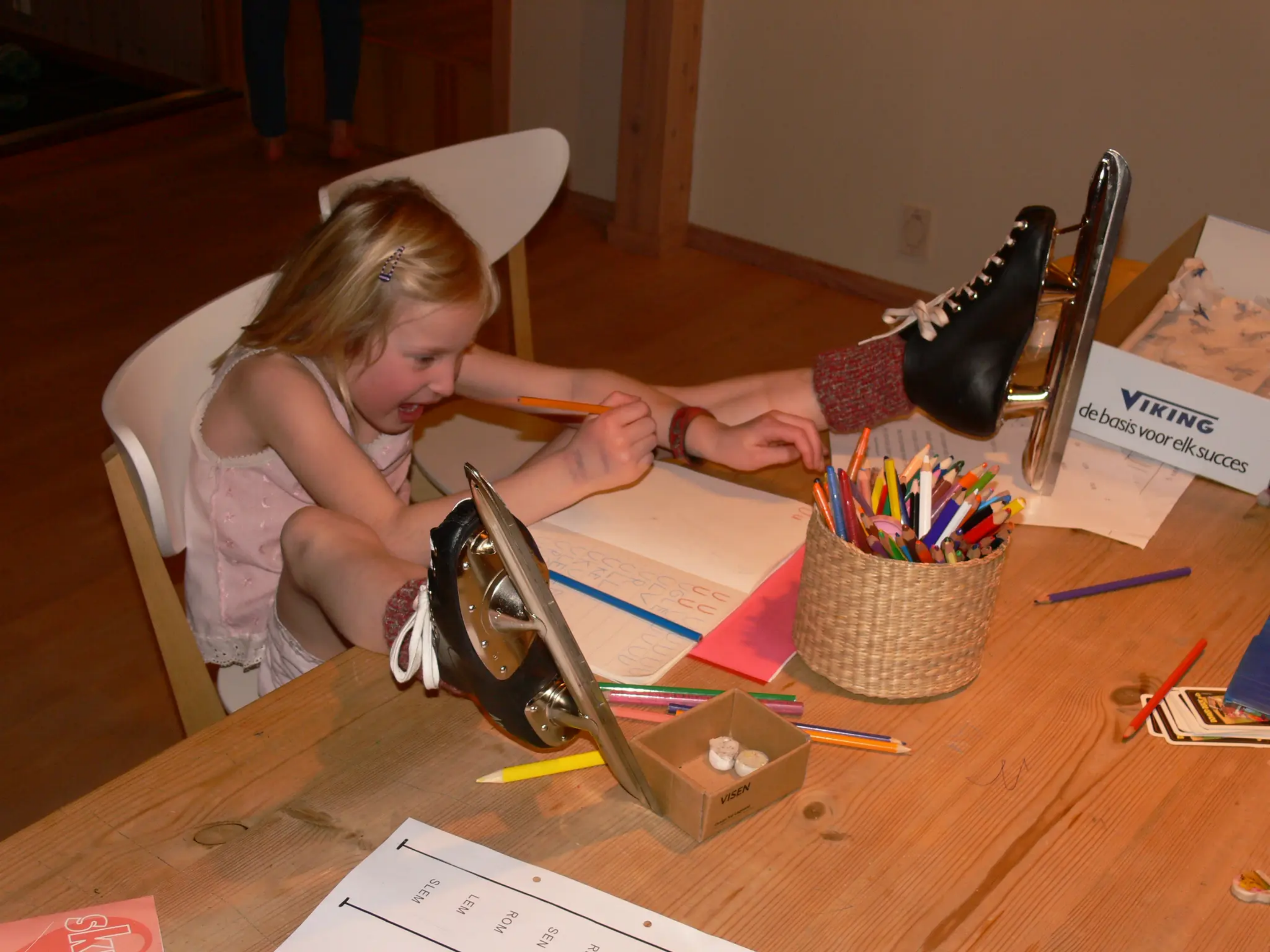}
\caption{
  How to win three Olympic medals:
  be good; be eager; do your homework; try different sports
  (e.g.~orienteering and skating); take your studies seriously;
  graciously and smilingly allow your competitors to win (occasionally).
  She also has the Zivilcourage to say (VG, 23/ii/2026),
  contrary to public perception and cultural expectation, 
  that she values her 2 silvers and 1 bronze more than
  the 1 gold she didn't win.}
\label{figure:ragne}
\end{figure}

But we are different (apparently), and we dare to be.
It's considered perfectly acceptable, even for the intellectuals
and for academics, to take a serious interest in the sociology,
history, and current nerdy details of sports,
and perhaps in the Winter Olympics sports and details in particular.
Read Knausgård's heart-warming essay on the secrets of speedskating,
read Thor Gotaas, read Eldar Høidal, read Rune Slagstad.
Perhaps the Svenska Akademien and the rest of the literary
world are utterly surprised and bewildered when they encounter
forty pages of speedskating details in the middle of
Dag Solstad's Nordisk Råds Litteraturpris winning novel. We're not.

\section*{Other Statistics-Olympic Perspectives}

I suppose I've managed to flag my above-average interest
in the Winter Olympics and the accompanying host of correlated
issues and themes. Here are just a few further statistics
related topics worthy of over-coffee conversations
(along with what in academia is called `further thought').


\smallskip
{\bf A.}
There is much talk of and posting of opinions and even
memes along the path of `With a US population sixty times
that of Norway, why do they still beat us',
supplemented by medals per capita tables, etc.
Ingar Kiplesund's Norwegian long jump record is 8,21 m,
as we know, which must correspond to a jump of length 499 m,
clearly more impressive than Powell's so-called world record
of a mere 8,95 m. 

Also, we're a Tiny Nation, with a healthy production of metal
per capita, but occasionally we're being Olympically
soundly outperformed by perhaps Liechtenstein, or Luxembourg,
in the medal-per-capita event. In 2018, Liechtenstein
had 1 medal for 37,531 inhabitans;
Norway, with her population being then 5,195,921 / 37,531 = 138.443
times bigger, captured only a modest 39 medals.

So what matters, taking sizes of populations seriously,
would be the {\it participation rates} -- how many
high-level skiers are there, in the US, in Sweden, Finland, Russia,
compared to the Norwegian numbers? Tables constructed
from such numbers (which are not accurate, but which could
be tentatively estimated) would have some interest.
I do think Norway would still win, though, even when
taking participation numbers into account.
See Hjort (2016a) for mathematics and statistics
pertaining to such themes, there discussed for the world
of top chess players; there are basically {\it two}
crucical factors at work, the participation $n$
and the variance $\sigma^2$, when forming basic predictions
of how clever the very most clever are. So different
nations would by tradition naturally foster higher $\sigma$
than others, even when taking participation into account. 

\smallskip
{\bf B.}
When discussing the naturally occurring questions
of why and how Norway apparently beat everyone else,
there are various paths to uncovering the contributing
factors. One is on statistical top level -- let's do
logistic regressions, for one nation at the time,
or multiple nations. So let's find data for BNP,
education level, Gini indexes of egalitarianism, 
military spending, corruption level,
budgets for sports, scores for laziness and cultural upbringing.
These covariates $x_1,\ldots,x_p$
can then be thrown to a statistical analyser, via
\beqn
p(x_1,\ldots,x_p)={\exp(\beta_0 + \beta_1x_1+\cdots+\beta_p x_p)
     \over 1 + \exp(\beta_0 + \beta_1x_1+\cdots+\beta_px_p)}
\eeqn 
modelling the percent of medals captured.
Such exercises would be feasible (just get the numbers)
and informative -- I actually tried this with $x$
taken to be the {\it democracy scores} for nations,
since I had these in my computer, via
the Cunen, Hjort, Nygård (2020) peace research project.
Surprise: it didn't matter at all; the relevant $\beta$
was close to zero.

These endeavours would clearly not be sufficient, however;
one must visit the lower levels, closer to the ground;
how sports are organised, with what levels of trainers
and coaches, with which mechanisms and incitements
to foster not merely high level outcomes, but the very highest.
Just as Norwegian teachers go to Finland to check out
the hows and whys of the reportedly very best school teachers
in Europe, before coming home to implement some of their ideas,
other sports administrators could come to Norway to check us out.
But, again, our tiny nation's tiny secrets are not really secrets.

\smallskip
{\bf C.}
This has clearly been the Worst Ever Winter Olympics for Norway.
The number of events where we didn't win a single medal has
never been so high. Checking golds only shows that we won
a mere 18 golds in 116 events, which means 98/116 = 84.5 \%
non-NOR-golds. In 1924 there were 12/16 = 75.0 \% non-NOR-golds.
This doesn't go quite as far as `everything can be proven by statistics'
or to Twain's front door of all the Lies, Damned Lies, and Statistics
we can hurl in the general Olympic direction of things,
but is a neutral reminder that we ought to concentrate our
statistical efforts (counting, tableing, analysing, relating,
comparing, modelling, predicting) on the more vital questions,
as driven by context.

\smallskip
{\bf D.}
I'm not particularly fond of the incessant focus on the
Medals and Only the Medals, and dislike the implied
and often explicit notion of The Horrible Painful For Ever Embarrassing
fourth place (`den sure fjerdeplassen' has about hundred thousand
google hits). There should be a special Norwegian Wood Medal
for the exceedingly honorable 4th place, and I humbly
ask the United Nations and the IOC to see to it that this is
institutionalised within the 2034 Utah Olympics. 
At any rate, we can count them!, and my possible favourite,
qua statistical long-term Olympic summary, is the notion of
Olympic Points: 7-5-4-3-2-1 for gold-silver-bronze-4-5-6.
Such a table can be seen and contemplated in Appendix A below,
with data found via Erik Helmersen's diligent efforts.
I also like the Fibonacci scale 13-8-5-3-2-1; people
in their homes can indeed try out different scales,
given the tables of outcomes,
and learn about the quite high Spearman correlation coefficients
across the different tables. 

\smallskip
{\bf E.}
Rather than counting medals for absolutely all of the
events (including what is sometimes called `circus events'),
one may of course form separate medal tables for the core
long-standing events. I've spent some nerdish half-hours
collecting and summarising data, Olympics for Olympics,
for the mulitiple events of long-track speedskating,
with medals for the different nations, through history 1924--2026,
men and ladies (where our erratic friend AI simply
couldn't do it). See Appendixes C and D.

\smallskip
{\bf F.}
High-level sophisticated statistics also famously
comes into play when the more ambitious nations select their
athletes, to maximise medal chances given the strict IOC
quotas. Everyone in the Netherlands know about {\it the matrix},
associated with complicated prediction algorithms. 

\begin{figure}[h]
 \centering
 \includegraphics[scale=0.33]{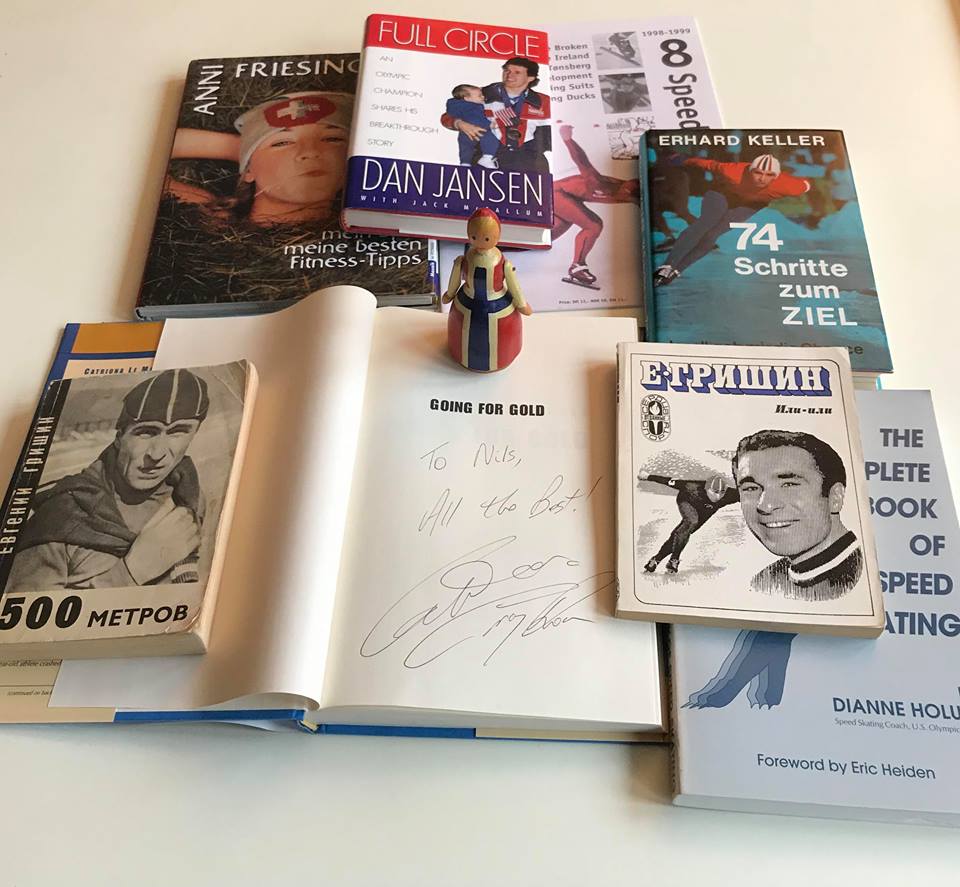} 
 \caption{
   A tiny subset of an educated household's
   shelves with speedskating history,
   here associated with various Olympic Gold Winners:
   Friesinger, Grishin, Heiden, Holum, Jansen,
   Keller, LeMay Doan, Postma, Rosa.}
 \label{figure:sprinterbooks}
 \end{figure}

\newpage 

\section*{Note} 

A plagiarism checker, should such algorithms take any interest,
will be able to find that some of the prose used in this note
indeed can be found in my FocuStat Blog Post from 2018,
written up after the Beijing Olympics. The self-plagiarism
component is moderate to low, though, as I've thorougly
dived into and organised more data, worked through
new perspectives, etc., and added more Olympic prose. 

\section*{Thanks} 


%

I'm grateful to all the 2922 participants in the 2026 Olympics
-- and, by backwards induction, to the participants of all
previous Winter Olympics (from the 2800 in Sochi 2014,
to the 1738 in Lillehammer and Hamar 1994, the 1094 in Innsbruck 1964,
the 494 in Oslo 1952, and all the 292 in Chamonix 1924).
I also appreciate that science journalists and
the Norwegian Broadcasting Corporation NRK apparently
read our FocuStat Blog Posts, with my 2 x 1000 m piece
leading to appearances in Abels Tårn (where I was properly starstrøkked)
and in Titan. For the 2026 Games specifically I applaud NRK's
efforts, to bring the events and also the grander perspectives 
to the people. 
Finally, a vigorous nod ought to be given
to the stimulating and variegated contributions from flocks
of eager people in the {\it Forum for Skøyehistorie} at Facebook,
where we watch and comment and discuss and shout and sigh
(and disagree) on a day-by-day basis during the bigger championships.

\section*{References} 

\begin{small}


\parindent0pt
\parskip3pt

Booth, M. (2014).
{\it The Almost Nearly Perfect People: The Truth About the Nordic
  Miracle.} Jonathan Cape, UK. 

Cunen, C., Hjort, N.L., Nygård, H.M. (2020). 
Statistical Sightings of Better Angels:
analysing the distribution of battle-deaths in interstate conflict
over time. Journal of Peace Research, 57, 221--234.

\font\cyr=wncyr9
{\cyr Grishin, E.R.} (1969). {\cyr 500 metrov.} Publisher:
{\cyr Molodaya gvardiya.}

Hjort, N.L. (1994).
Should the Olympic speedskaters race the 500 m twice?
Statistical Research Report, Department of Mathematics,
University of Oslo. (Yes, this report changed the Olympics.)

Hjort, N.L. (2017).
But some are more equal than others. FocuStat Blog Post.

Hjort, N.L. (2017).
The semifinals factor for skiing fast in the finals.
FocuStat Blog Post.

Hjort, N.L. (2018).
The Best Metal-Grabbing Games Ever. FocuStat Blog Post.

Hjort, N.L. (2018).
One Thousand is Unfair, Two Thousand is Fair. FocuStat Blog Post.

Hjort, N.L. (2025).
{\it Skøyteløpenes uutholdelige letthet} (the unbearable lightness
of speedskating).
Essay, Klassekampen (yes, Class Struggle),
15/xii/2025 (check the FocuStat site for a facsimile,
with a list of the eleven hidden literary references, and more). 

Hjort, N.L. (2026a).
Anyone for chess? Analysing chess ratings above high thresholds.
FocuStat Note, January 2026;
on arXiv, arxiv.org/abs/2602.04353.

Hjort, N.L. (2026b).
Six-Minute Man Sander Eitrem 5:58.52 -- 
first man below the 6:00.00 barrier
FocuStat Note, February 2026;
on arXiv, arxiv.org/abs/2602.03274. 

Hjort, N.L., Stoltenberg, E.Aa. (2026).
{\it Statistical Inference: 600 Exercises, 100 Stories.}
Cambridge University Press (to appear).

Høidal, E. (2012).
{\it To indre og vekk me'n.} Cappelen-Damm, Oso.

Jacobsen, R. (1991). {\it D\aa semikler.}
In {\it Seierherrene}, Cappelen, Oslo. 

Knausgård, K.O. (2018).
{\it The Hidden Drama of Speedskating}.
Essay, The New York Times, 1-Feb-2018.


Løchen, J.F. (1952).
{\it Vi viste verden vinterveien.}
Gyldendal, 1952. 

Pechstein, C. (1990).
{\it Von Gold und Blut: Mein Leben zwischen Olymp und Hölle.}
Schwarzkopf \& Schwarzkopf. 

Rosa, D. and Hjort, N.L. (1999).
{\it Who Won? The Annual Speedskating Race of the Burg of Ducks.}
Speedskating World. (Also included in Don Rosa Collected Works, 2011;
translated to Danish, Finnish, German, Dutch, Norwegian, Swedish.) 

Schwanitz, D. (1999).
{\it Alles, was man wissen muß.} Eichborn, Frankfurt am Main.

Schweder, T., Hjort, N.L. (2016).
{\it Confidence, Likelihood, Probability:
  Statistical Inference with Confidence Distributions.}

Solstad, D. (1989). {\it Roman 1987.} Oktober, Sandefjord.

Slagstad, R. (2008). {\it (Sporten) En idéhistorisk studie}. Pax, Oslo.

Thunberg, C. (1947). {\it Alene mot hele Norge.}
Oversatt av Finn Amundsen. Oslo: Ekko Forlag. 

Vold, J.E. (1994). {\it En sirkel is.} Gyldendal, Oslo. 

\end{small}


\newpage 

\section*{Appendix A: Medals \& Olympic Points} 

Here I list nations, sorted with respect to medals won,
giving also the Olympic Points to the right,
using the historically informed 7-5-4-3-2-1 scale,
along with the consequent spearmanny well correlated rank list.
AIN is the {\it athlètes individuels neutres} category,
all to be admired for competing under exceedingly
difficult circumstances, both at home and here. 

\begin{small} 
\begin{verbatim}

        gold silver bronze total     OP   ranking 
 1 NOR	  18    12    11     41     295.0    1
 2 USA	  12    12     9     33     254.5    2
 3 ITA	  10     6    14     30     220.0    3
 4 GER	   8    10     8     26     192.0    4
 5 JPN	   5     7    12     24     145.5    9 
 6 FRA	   8     9     6     23     178.5    5
 7 SUI	   6     9     8     23     161.0    6
 8 CAN	   5     7     9     21     152.0    6 
 9 NED	  10     7     3     20     146.0    8
10 SWE	   8     6     4     18     122.5   11
11 AUT	   5     8     5     18     131.0   10
12 CHN	   5     4     6     15     113.0   12
13 KOR	   3     4     3     10      65.0   13
14 AUS	   3     2     1      6      48.0   15
15 FIN	   0     1     5      6      44.0   16 
16 GBR	   3     1     1      5      53.0   14
17 CZE	   2     2     1      5      38.0   17
18 SLO	   2     1     1      4      29.0   18
19 POL	   0     3     1      4      28.0   19
20 ESP	   1     0     2      3      17.0   21
21 NZL	   0     2     1      3      16.0   22  
22 LAT	   0     1     1      2      22.0   20
23 BUL	   0     0     2      2      11.0   25
24 BRA	   1     0     0      1       7.0   28
   KAZ	   1     0     0      1       8.0   26
26 DEN	   0     1     0      1       5.0   30
   EST	   0     1     0      1       6.0   29
   GEO	   0     1     0      1       8.0   26
   AIN    0     1     0      1      13.0   23
29 BEL	   0     0     1      1      12.0   24  
   SVK                               3.0   31
   HUN                               3.0   31
   ROM                               2.0   33
   UKR                               2.0   33
   LIT                               1.0   35

total:   116    118    115  349          2552

\end{verbatim}
\end{small}


\newpage

\section*{Appendix B: The Norwegian Medal Table, 1924 to 2026} 

The table gives for each Olympics the number of events;
then the metal count (with total);
followed by the number of nations taking part; 
and the percent of medals won by Norwegian athletes.

\begin{small}
\begin{verbatim} 
                        ev   g  s  b  tot  nations percent
Chamonix         1924   16   4  7  6   17     16     35.4
St Moritz        1928   14   6  4  5   15     25     35.7
Lake Placid      1932   14   3  4  3   10     17     23.8
Garmisch         1936   17   7  5  3   15     28     29.4
St Moritz        1948   22   4  3  3   10     28     15.2
Oslo             1952   22   7  3  6   16     30     24.2
Cortina          1956   24   2  1  1    4     32      5.6
Squaw Valley     1960   27   3  3  0    6     30      7.4
Innsbruck        1964   34   3  6  6   15     36     14.7
Grenoble         1968   35   6  6  2   14     37     13.3
Sapporo          1972   35   2  5  5   12     35     11.4
Innsbruck        1976   37   3  3  1    7     37      6.3
Lake Placid      1980   38   1  3  6   10     37      8.8
Sarajevo         1984   39   3  2  4    9     49      7.7
Calgary          1988   46   0  3  2    5     57      3.6
Albertville      1992   57   9  6  5   20     64     11.7
Lillehammer      1994   61  10 11  5   26     67     14.2
Nagano           1998   68  10 10  6   25     72     12.3
Salt Lake Placid 2002   78  13  5  7   25     78     10.7
Torino           2006   84   2  8  9   19     80      7.5
Vancouver        2010   86   9  8  6   23     82      8.9
Sochi            2014   98  11  5 10   26     88      8.8
PyeongChang      2018  102  14 14 11   39     92     12.7
Beijing          2022  109  16  8 13   37     91     11.3  
Milano Cortina   2026  116  18 12 11   41     92     11.8
\end{verbatim} 
\end{small}

\vspace{-0.2cm}

\section*{Appendix C: medal counts for speedskating, 1924-2026}

I now focus on the iconic classic events of long-track
speedskating, through Olympic history. The table on the
next pages gives {\it all medals won},
for men and ladies (the term used by the ISU over hundred years,
but now apparently abandoned, for the more prosaic `women').
Readers are advised to print out my report and
then to pencil in vertical bars. 
 
Medals have been won by {\it 19 nations} for the men
and {\it 17 nations} for the ladies. As we recall,
Han Pil-hwa of North Korea won a 3k silver in 1964, and for
statistical simplicity I've pushed her under KOR in
this table (though see Appendix D). 
I've also lumped together BRD and DDR with their GER.

The political history of states is reflected here: CCCP took part
from 1956 to 1988, after which RUS and BLR and KAZ after
an interregnum period came into the tables.
Medals have exploded, if not in sizes then in numbers:
for 1924--1956, there were {\it 4 events} (12 medals to fight for);
then ladies were let in in 1960,
and with the addition of {\it team pursuit} since 2006
and {\it mass start} since 2018, there are now
an astounding {\it 14 events} (42 medals to fight for, each Olympics).  

\newpage

\begin{footnotesize}
\begin{verbatim}
     medal table for the speedskating men: 
     nor ned usa fin swe can aut sov ger jpn kor rus blr bel ita pol chn cze den 
     85  86  45  17  18  20  3   28  10  12  16  7   2   3   8   3   5   2   1
1924  6       1   6
1928  6       1   4
1932  2       5           5
1936  6       1   4           1
1948  6       2   2   3

1952  6   3   2       1   1
1956  2           1   2           7
1960  3   1   1       1           6
1964  7   1   1       1           3
1968  4   4   1       2               1

1972  4   4           2           1   1
1976  4   5   2                   4
1980  6   2   5           1       2
1984  1               2   3       4   2   1
1988      4   1       2       2   2   3   1

1992  5   4                           2   3   1
1994  5   4   1                               1   3   1
1998  1   9               2               2               1
2002  2   6   5                       1   1   
2006      5   7           1                   1   1           3

2010  1   4   4           1               2   4   2           
2014     13               2               1                       2
2018  4   7               2                   5           1   1       1
2022  3   5   1       2   1               1   4   1       1   1       1
2026  1   5   4           1                                   3   1   3   2   1
     medal table for the speedskating women:
     nor ned usa fin swe can aut sov ger jpn kor rus ita pol chn cze kaz
     5   60  33  2   0   27  3   30  61  17  5   9   4   4   7   7   1
1960          1   1               6   2                   2
1964          2                  10           1
1968      5   4   1               2          

1972      5   4                   2   1
1976   1      4           1       5   1
1980   1  2   3                   2   4
1984                              3   9
1988      3   2                      10

1992          2               1       9   1                   2
1994          2           1   2       5   1       2           1
1998      2   2           3           6   1                           1
2002      2   3           3           7
2006      4               7           3           2           2

2010      3               4           4   1   1           1   1   3 
2014     10                                   1   3       1   1   2
2018      9   1                           6   2   1               2
2022      7   2           4               4       1   2           1
2026  3   8   1           4               3           2
  
\end{verbatim} 
\end{footnotesize}

\newpage

\section*{Appendix D: Speedskating medals, men + ladies}

Here I again single out long-track speedskating for scrutiny,
with accumulated numbers of medals, through Olympic history
1924--2026.

When Jorrit Bergsma won the mass start,
with his characteristic haircut, amazingly securing his fourth medal,
he also ended Hundred Years of Norwegian Male Dominance;
NED 86 $>$ NOR 85. Among other notable aspects here are these.
(i) Egalitarian Norway has not really minded enough about
the women, who had only a very modest 2 before Wiklund
now earned 3 more.
(ii) Sweden, where art thou? Zero ladies medals, and
even though your men are occasionally fabulous (particularly
in the 10k), we need more of you.
(iii) I've lumped together BRD, DDR, GER. Carefully
looking through 25 wikipedia articles (which I did to
manually harvest the data for Appendixes C and D),
one is reminded of the supreme dominance of the DDR
skateresses, over many years.
(iv) We do hope you became and remain a national
hero of the DPRK, Han Pil-hwa, with your silver from 1964,
but we long for your comrades: come and take part. 
(v) Good for you, Viktor Thorup, impressive silver;
next time bring also your wife, this year's European
Champion in that event.
(vi) So it's 3 + 0 = 3 for BEL, and perhaps you abstained
from winning the 5k, Sandrine Tas, just out of politeness
to your NED pairmate. 

{{\baselineskip12.0pt 
\begin{center}    
\begin{verbatim}
      NED  86 + 60 = 146
      NOR  85 +  5 =  90
      USA  45 + 33 =  78
      GER  10 + 61 =  71 
      SOV  28 + 30 =  58

      CAN  20 + 27 =  47
      JPN  12 + 17 =  29
      KOR  16 +  4 =  20
      FIN  17 +  2 =  19
      SWE  18 +  0 =  18

      RUS   7 +  9 =  16
      ITA   8 +  4 =  12
      CHN   5 +  7 =  12
      CZE   2 +  7 =   9
      POL   3 +  4 =   7

      AUT   3 +  3 =   6 
      BEL   3 +  0 =   3
      BLR   2 +  0 =   2
      NKR   0 +  1 =   1
      KAZ   0 +  1 =   1
      DEN   1 +  0 =   1
\end{verbatim} 
\end{center}    
}}

\newpage 

\section*{Why the Olympics}

Why peace, not war? The Olympics is a Peace Project,
perhaps not since Ancient Greece 776 b.C.~but in principle at least
since 1896 a.C. There are setbacks and disappointments
and boycotts and polarisation and politics and
sanctions against innocent fine athletes,
who perhaps have been born with the wrong passport. 

It is hopeless and we don't give in,
as Norway's national poet Jan Erik Vold writes,
an author also of speedskating poetry, a two-word
combination making perfect sense in Norway if not
in Andorra or Zimbabwe -- but we're getting there.
We need more time.

We also need the cultural continuity, across and beyond
the four-year cycles. Yes, we do need snow and ice
and fabulous athletes who apparently need their medals.
By extension the world also needs {\it recruitment},
in the broadest sense. Perhaps more important than
young clever people's dreams of {\it becoming} Olympians
is other young clever boys and girls and their dreams
{\it about} Olympics.
Hello, young nerdy boys and girls, go for it,
immerse yourselves, do watching and noting and comparisons
and tables and figures and of course follow the laptimes
of speedskating -- and your lives will be meaningful and good.
Yes, admittedly, even Solstad refers to these matters as
`empty feelings', but they build character and metaphors
and literary instincts and poetry and fascination
and spell-binding and finess the speculative instincts we're born with
(which is helpful when you attempt to fall in love
or do your homework). 
And, indeed, albeit slowly, to friendships and peace. 

\bigskip

\begin{figure}[h]
 \centering
 \includegraphics[scale=0.60]{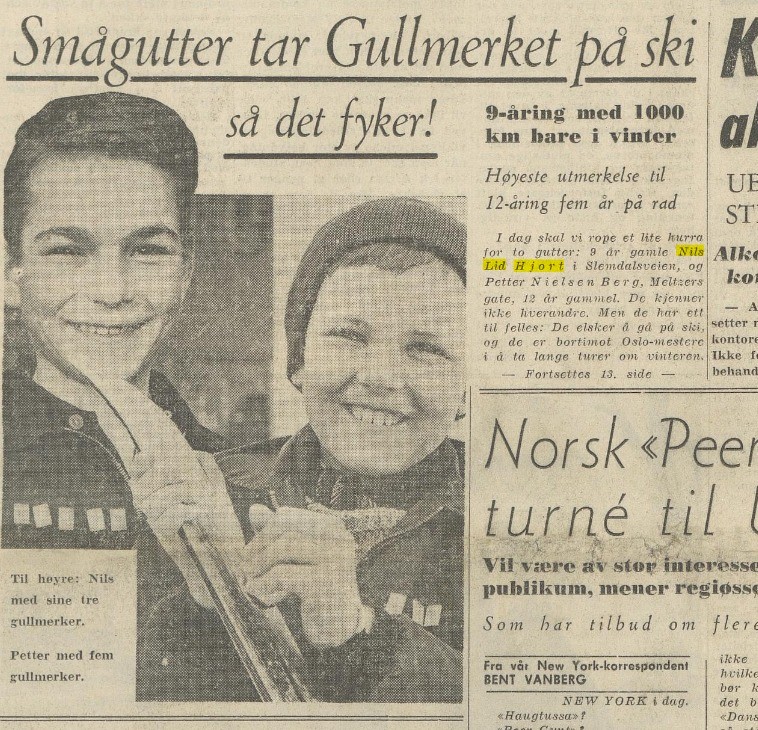} 
 \caption{
   9-åring med 1000 km på ski bare i vinter!
   Front page of Dagbladet (once upon a time).}
 \label{figure:queen}
 \end{figure}

\eject

\end{document}